\documentclass[aps,prev,preprintnumbers,floatfix,nofootinbib]{revtex4}
\pdfoutput=1

\usepackage{amsmath}
\usepackage{amsfonts}
\usepackage{amssymb}
\usepackage{graphicx}
\usepackage{bm}
\usepackage{times}
\usepackage{hyperref}
\usepackage{slashed}
\usepackage{color}

\begin{document}

\title{Brazilian Report on Safeguards Application of Reactor Neutrinos}

\author{$^1$E. Kemp\footnote{Corresponding author: kemp@ifi.unicamp.br}}
\author{$^2$J. A. M. Alfonzo}
\author{$^3$J. C. Anjos}
\author{$^3$G. Cernicchiaro}
\author{$^4$P. Chimenti}
\author{$^5$I. A. Costa}
\author{$^2$P. C. M. A. Farias}
\author{$^5$A. Fernandes Jr.}
\author{$^6$G. P. Guedes}
\author{$^1$L. F. G. Gonzalez}
\author{$^3$H. P. Lima Jr.}
\author{$^5$A. S. Lopes Jr.}
\author{$^2$J. Marcelo}
\author{$^5$M. L. Migliorini}
\author{$^5$R. A. N\'obrega}
\author{$^2$I. M. Pepe}
\author{$^2$D. B. S. Ribeiro}
\author{$^1$W. V. Santos}
\author{$^2$D. M. Souza}
\author{$^2$L. R. Teixeira}
\author{$^4$A. M. Trzeciak}

\affiliation{$^1$Universidade Estadual de Campinas, Campinas, SP, Brazil}
\affiliation{$^2$Universidade Federal da Bahia, Salvador, BA, Brazil}
\affiliation{$^3$Centro Brasileiro de Pesquisas F\'{i}sicas, Rio de Janeiro, RJ, Brazil}
\affiliation{$^4$Universidade Estadual de Londrina, Londrina, PR, Brazil}
\affiliation{$^5$Universidade Federal de Juiz de Fora, Juiz de Fora, MG, Brazil}
\affiliation{$^6$Universidade Estadual de Feira de Santana, Feira de Santana, BA, Brazil}

\begin{abstract}
The Neutrinos Angra Experiment is a water-based Cherenkov detector located in the Angra dos Reis nuclear power plant. The experiment has completed a major step by finishing the commissioning of the detector and the data acquisition system at the experimental site. The experiment  was designed to detect the electron antineutrinos produced by the nuclear reactor with the main purpose to demonstrate the feasibility of monitoring the reactor activity using an antineutrino detector. This effort is within the context of the International Atomic Energy Agency (IAEA) program to identify potential and novel technologies that can be applied for non-proliferation safeguards. Challenges, such as operating at the surface, therefore with huge noise rates, and the need to build very sensitive but small-scale detectors, make the Angra experiment an excellent platform for developing the application itself, as well as acquiring expertise in new technologies and analysis methods.  In this report, we describe the main detector features and the electronics chain (front-end and data acquisition). We also report preliminary physics results obtained from the commissioning phase data.  Finally, we address conclusions regarding the future perspectives to keep this program active, due to its importance in the insertion of Latin-American scientists and engineers in a world-scale cutting edge scientific program.

\end{abstract}

\maketitle

\section{Introduction and Motivation}
Nuclear reactors have been crucial to experimental neutrino physics as they are copious man-made sources of neutrinos. In addition to having made possible the confirmation of the neutrino hypothesis in 1956 \cite{cowan}, other experiments have been built since then inside such facilities. 
We should mention the crucial contribution of reactor neutrinos experiments in the precise knowledge of neutrino oscillation parameters \cite{kamland, dchooz, dbay, reno}. 


The neutrino emission from nuclear reactors is closely related to the fission of heavy nuclei taking place inside the reactor: each fission contributes with a well-known fraction of the total released energy and is followed by the emission of neutrinos.  By using a neutrino detector to monitor the flux, one can estimate the fission's rate and thus the energy released by the reactor, i.e., the neutrino flux is directly proportional to the power of the reactor.

The proposal to use neutrinos for remote monitoring of the thermal power of nuclear reactors was first considered in the mid-1970s \cite{Mik,BoroMik}. One of the first demonstration experiments was performed in a neutrino laboratory located in the nuclear power plant at Rovno - Ukraine. The relationship between neutrino counting rate and the reactor activity was clearly shown, so the neutrino radiation could be, in principle, used for such purposes \cite{Rovno1,Rovno2}.

This scenario described above opens up solid perspectives for the use of neutrinos as reliable probes of the physical processes in which they participate. Thus, a neutrino detector can monitor parameters related to the activity of nuclear reactors that are crucial for checking items of the non-proliferation safeguards dictated by the International Atomic Energy Agency (IAEA).

An important point in the use of neutrino detectors for nuclear reactor monitoring, especially regarding the verification of safeguards, is the possibility to remotely check the reactor activity, avoiding the need for operation and intrusion in the containment area or other restricted access areas in the nuclear power plant. Summaries of the worldwide effort on the use of antineutrinos detectors for safeguards purposes can be found in \cite{safeguards, korea} and references therein. 

A neutrino experiment at the Brazilian nuclear power plant in Angra dos Reis was initially considered for studies of neutrino oscillations \cite{angra-1st}. After the alignment of the groups around the three proposals that would be implemented (Double Chooz, RENO and Daya Bay), the Brazilian group decided that there was a great opportunity to perform an experiment in the premises of the plant, even if of smaller dimensions and with another purpose, in this case the verification of non-proliferation safeguards. This is the origin of the Neutrinos Angra ($\nu$-Angra) experiment, discussed in this report.

The Angra-II reactor, in steady-state operation, provides a neutrino flux estimated as $1.21\times10^{20}~s^{-1}$ \cite{connie}. The electron antineutrino flux can be used to perform a non-invasive monitoring of the reactor activity as well as to estimate the thermal power produced at the reactor core. 

The long term goal of the $\nu$-Angra experiment is to develop a reliable and cost-effective technology to routinely monitor the nuclear reactor power and possibly the neutrino spectral evolution \cite{hubber}, which can unveal the composition of burned fuel, with special interest to the fraction of plutonium, the main target for diversion. The stability of the data acquisition is a fundamental step in this direction. This goal is well aligned with the safeguards and non-proliferation demands of the IAEA \cite{IAEA-anutech}. In the current commissioning phase, we have fully validated the electronics, which is operating within the desired stability to run for years of data taking.

A common challenge for all neutrino experiments, using reactors or not, is the background radiation -  mainly neutrons and cosmic muons - that can mimic signals with similar features of those expected from neutrino detection. To suppress the background level, the usual solution is to install the detectors in large underground caverns \cite{dchooz} using the rock and soil overburden as a natural shield. Several measurements have shown that the vertical muon intensity can be reduced $10^{4}$ times for $10^{2}~hg~cm^{-2}$ depth of rock \cite{grieder}.

The $\nu$-Angra detector is assembled on the surface, one of the points of the agreement with the plant operator to conduct the research inside the Angra dos Reis nuclear complex. In consequence, the main challenge (if not the biggest)  is to achieve a Signal-To-Noise Ratio high enough in order to make possible to monitor the burning process of the Agra-II nuclear reactor\footnote{The Angra dos Reis nuclear complex has 2 reactors in operation and a third under construction. Angra-II is the most powerfull currently in operation ($\sim$ 4 GW). The detector is placed near the containment dome external wall.}.
The $\nu$-Angra experiment is taking data in stable manner since late 2018 \cite{AngraCommissioning}. The detector was successfully commissioned in between September 2017 and the second half of 2018. In this report we briefly describe the detector elements and functional features, the main physical results of the preliminary data analysis, and finally the conclusions and perspectives for the experiment.

\section{The $\nu$-Angra Detector}
\subsection{Detector Overview}

\begin{figure}
	\centering
	\includegraphics[scale=0.4]{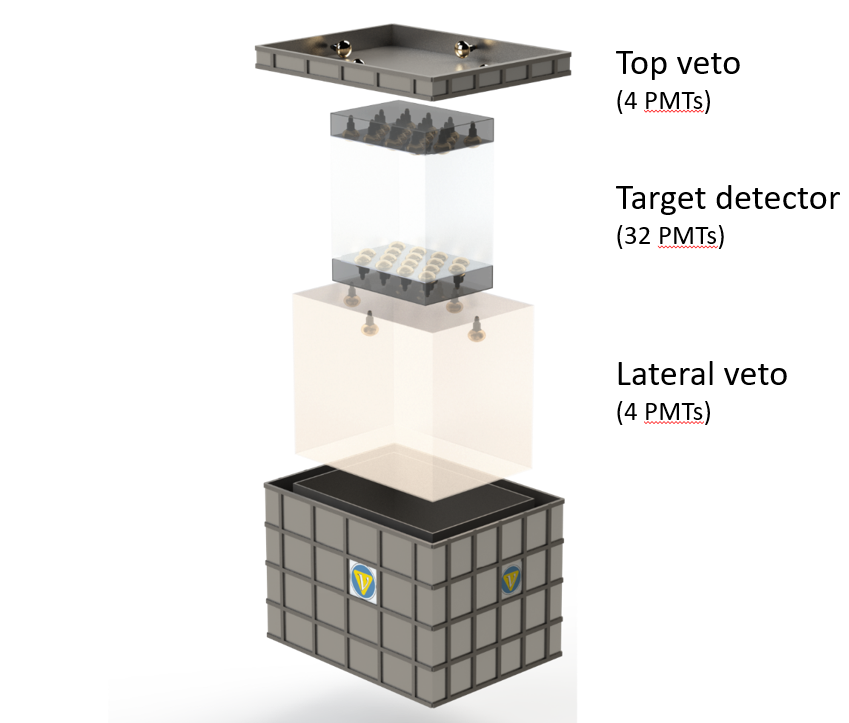}\\
	\caption{Exploded view of the $\nu$-Angra detector.}
	\label{fig:detector}
\end{figure}

The $\nu$-Angra is a water Cherenkov detector. 
The electron antineutrino flux generated by the reactor can be used to perform a non-invasive monitoring of the reactor activity as well as to estimate the thermal power produced at the reactor core. For the 1~ton Target detector, the expected rate is around 5000 events per day for a distance of 25~m from the reactor core, enabling the experiment to investigate the potential of antineutrino detection for safeguards applications. The antineutrino interactions are recognized from inverse-$\beta$ decay events, taking place inside the Target detector, a cubic volume filled with 1340.28 l of $GdCl_{3}$ doped water (0.2~\%). The water volume is contained by a 0.90~m$~\times~$1.46~m$~\times~$1.02~m (height-length-width) plastic tank surrounded by 32 eight-inch photomultiplier tubes (PMTs), with 16 installed on the bottom and 16 on the top of the volume, as is shown in figure \ref{fig:detector}. The six internal faces of the target volume are covered by sheets of a diffuse reflector membrane (0.5~mm DRP), which provides reflectance greater than 99.0~\% for 400~nm wavelength light.

The vessel on the top of figure \ref{fig:detector} is the Top VETO detector, designed to detect and generate a veto pulse whenever a cosmic muon crosses the volume. Four PMTs, of the same model as the one in the Target detector, are fixed on the middle of each side, pointing to the center of its volume. The Top VETO tank is filled with pure water, has dimensions 0.28~m$~\times~$2.66~m$~\times~$2.02~m (h-l-w) and has all the internal faces covered with the Tyvek\footnote{Trademark of DuPont.} reflector material (providing reflectance greater than 97.0~\%). Surrounding the Target detector, there is a first water layer with 12.00~cm thickness. In this part of the detector - called here the Lateral VETO - there are four PMTs fixed in the middle of each side on the top with the photocatode pointing to the bottom. The Lateral VETO corresponds to the third part - from top to bottom - in figure \ref{fig:detector}. The Top and the Lateral VETO layers (each one covered by four photomultipliers), and a dedicated circuitry, form the VETO system, which is responsible for detecting cosmic ray particles that might hit the Target detector volume. Whenever two or more PMTs fire at the same time, a veto window is generated blocking any trigger signal that could eventually be generated due to hits in the Target detector. 

Finally, the outermost part of the detector, which surrounds the Lateral VETO, is 14.50 cm thick on two opposite sides and 22.5 cm on the other two sides. It is completely filled with pure water and used as shield against background neutrons.       

\subsection{Readout Electronics}
\begin{figure}[ht!]
	\centering
	\includegraphics[scale=1.0]{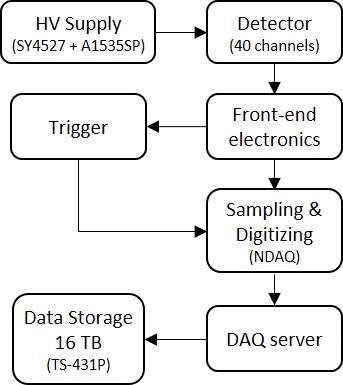}\\
	\caption{Overview of the Readout Electronics installed on the nuclear power plant.}
	\label{fig:daq_overview}
\end{figure}

In addition to the detector itself, a complete data acquisition system has been designed and integrated for the experiment in order to perform tasks, such as: biasing of the PMTs (High Voltage power supply), amplification of PMT output signals (Front-End electronics), sampling and digitization of the signals (NDAQ electronics), online events selection (Trigger system) and local data storage (commercial \emph{Network Attached Storage}). An overview of the Readout Electronics with those systems is the diagram shown in the figure \ref{fig:daq_overview}.

The high voltage system and the local computing infrastructure - DAQ (Data Acquisition System) servers and local data storage - are commercial off-the-shelf parts in the readout electronics. The High Voltage Power Supply used to bias the 40 PMTs is a commercial mainframe-based system from CAEN. The HV system is remotely controlled through an Ethernet connection. The local data storage is a \emph{network attached storage} unit. Data read from the detector is continuously written to those disks and subsequently transferred to two larger and permanent data servers: the primary one installed at CBPF, in Rio de Janeiro, and a mirror server installed at Unicamp, in Campinas.

\subsubsection{Front-end electronics}
\begin{figure}[ht!]
	\centering
	\includegraphics[scale=0.33]{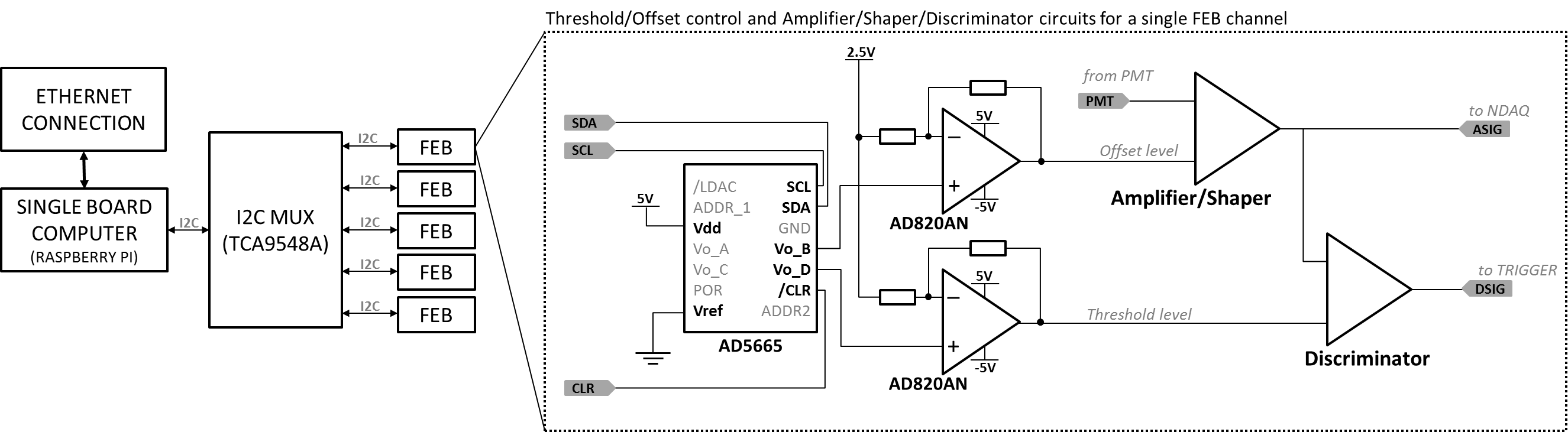}\\
	\caption{Front-end electronics channel.}
	\label{fig:fee_schematics}
\end{figure}

In order to condition the PMTs output signals to the digitizing electronics (NDAQ modules), and to inform which channels have been fired to the trigger system, a custom front-end circuitry has been developed \cite{fee_1}. Five front-end boards (FEB) are used in the experiment, each one with eight independent channels. A channel is composed of a four-stage amplification/shaper circuit, a discriminator circuit and a control system. The former prepares the signal to the digitizer, the output of the second is delivered to the trigger system and the latter allows remote adjustment of the offset of the analog signal and the threshold of the discriminator circuit according to the experiment requirements. Therefore, as shown in figure \ref{fig:fee_schematics}, each channel offers two output signals: an analog (ASIG) and a discriminated (DSIG) signal.

The control system, also shown in figure \ref{fig:fee_schematics}, is based on controllable DAC chip, placed on the FEB, and a single-board computer (RaspberryPI) connected to the network to allow accessing all the DAC (Digital-to-Analog Converter) chips remotely.
The design of the front-end circuit considered the region of few tens of photoelectrons where the amplification circuit is linear, the electronic noise is less than 0.1 photoelectron and the threshold resolution is approximately 0.05 photoelectron.


\subsubsection{Digitizer}
The detector signals, amplified and shaped by the front-end electronics, are sent to a VME-based system able to sample and digitize the signals whenever it receives a trigger pulse. The NDAQ board was specifically designed for the experiment \cite{ndaq}. A VME single board computer works as the readout processor (ROP) of five NDAQ boards inserted in the VME crate. The ROP controls and reads the NDAQ cards during data taking. A commercial  fan-in fan-out module is used to distribute the digital trigger pulse coming from the Trigger system to the NDAQ modules.

\begin{figure}[ht!]
	\centering
	\includegraphics[scale=0.35]{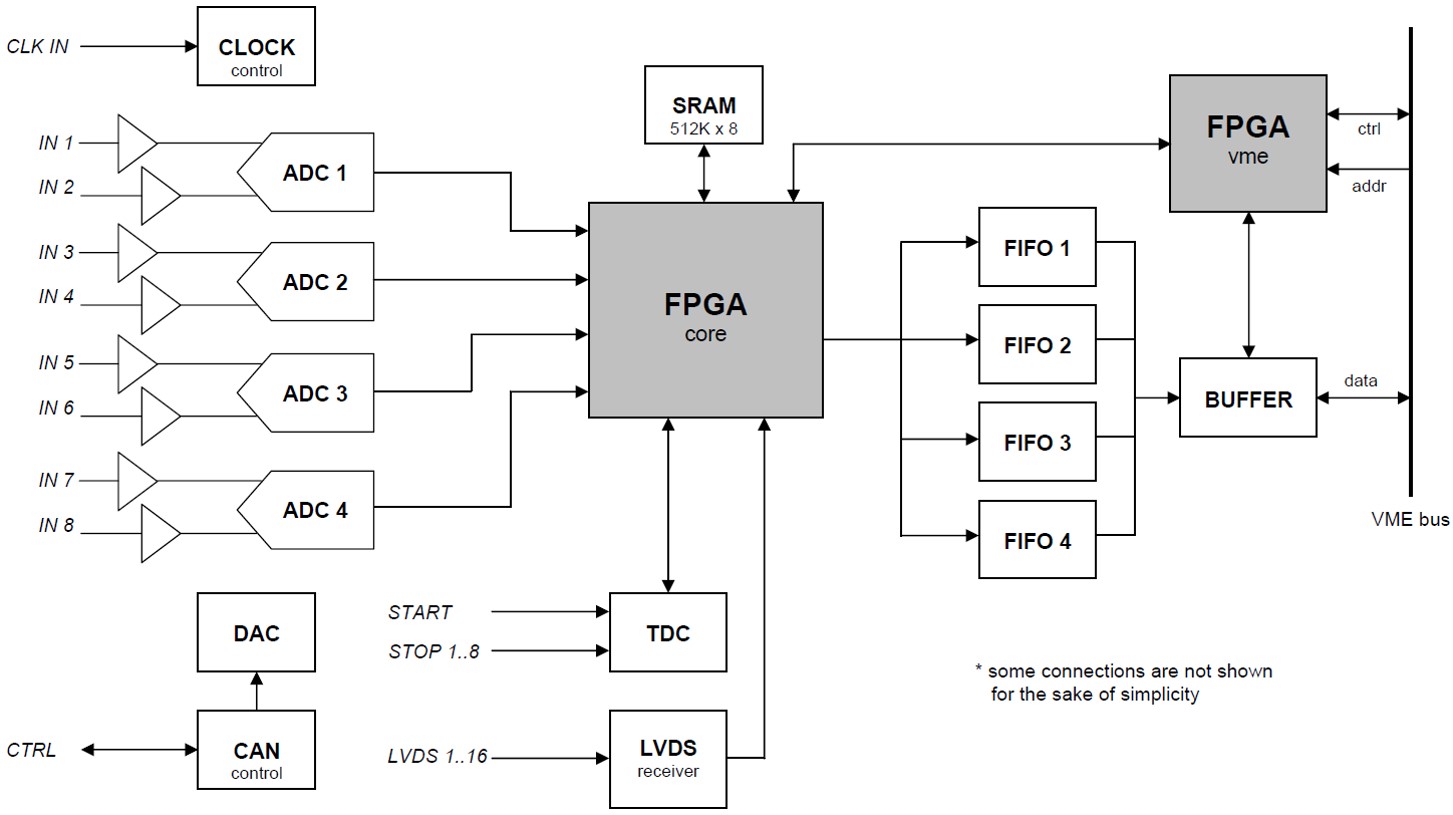}\\
	\caption{NDAQ data acquisition module design.}
	\label{fig:ndaq_diag}
\end{figure}

Each NDAQ module has eight Analog-to-Digital Converter (ADC) channels working at 125~MHz sampling rate (10~bits resolution), as shown in figure \ref{fig:ndaq_diag}. The ADC output samples are sent to a Field Programmable Gate Array (FPGA) which controls the data flow to the two FIFO memories connected in series. 

\subsubsection{Trigger}
\begin{figure}[ht!]
	\centering
	\includegraphics[scale=0.8]{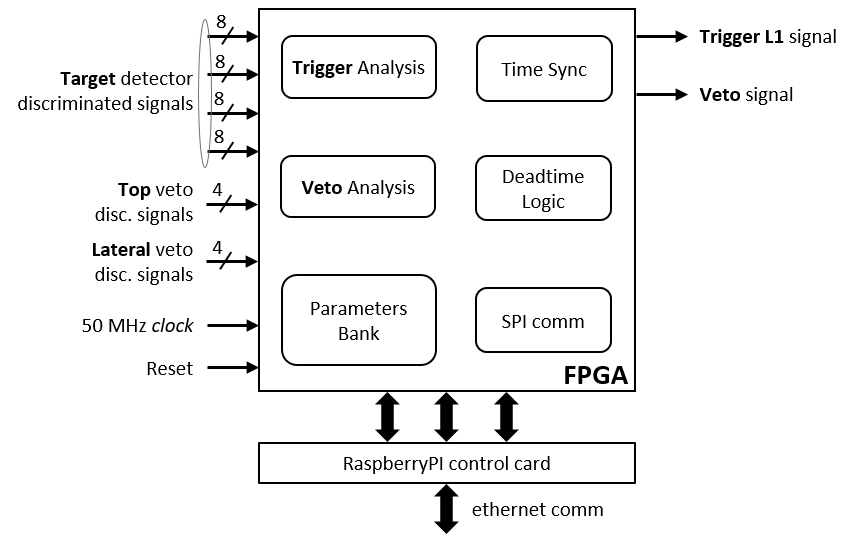}\\
	\caption{Trigger electronics overview.}
	\label{fig:trigger_diag}
\end{figure}

In the occurrence of an event, the front-end output signals are digitized by the NDAQ modules and sent to on-board FIFO memories waiting for a trigger decision. If the event is selected by the Trigger System \cite{trigger}, shown in figure \ref{fig:trigger_diag}, the corresponding data is transferred to the experiment data storage unit for future analysis. For a fast trigger decision, the selection algorithm was developed to be implemented in a dedicated FPGA. For remote configuration and upgrade of its firmware, a RaspberryPI card with Ethernet connection has been integrated to the FPGA circuit.

The Trigger board receives as input the 40 discriminated signals from the Target, the Lateral Veto, and the Top Veto. The trigger signal is formed whenever a minimum of PMT units have been fired in the target within a prescaled time window, and is blocked when a veto condition was found. 

Different trigger configurations are available and can be configured by remote access to a RaspberryPI control card. 

\section{Commisioning data: preliminary results} 
\subsection{Operations and running conditions}

Data collected during the commissioning phase were analyzed to check the proper functionality of the whole system. Main results are:
\begin{itemize}
    \item stability: we achieved the stability of trigger rate and operational conditions required for long term operations (years of data taking).
    \item FEE and DAQ: firmware, hardware and integration of all electronics subsystems run accordingly to the experiment requirements.
    \item remote operations: we checked the feasibility of remote operations. Few on-site interventions are needed only in extreme cases.
\end{itemize}

\subsection{Preliminary physics analysis}

\begin{figure}[ht!]
	\centering
	\includegraphics[scale=0.3]{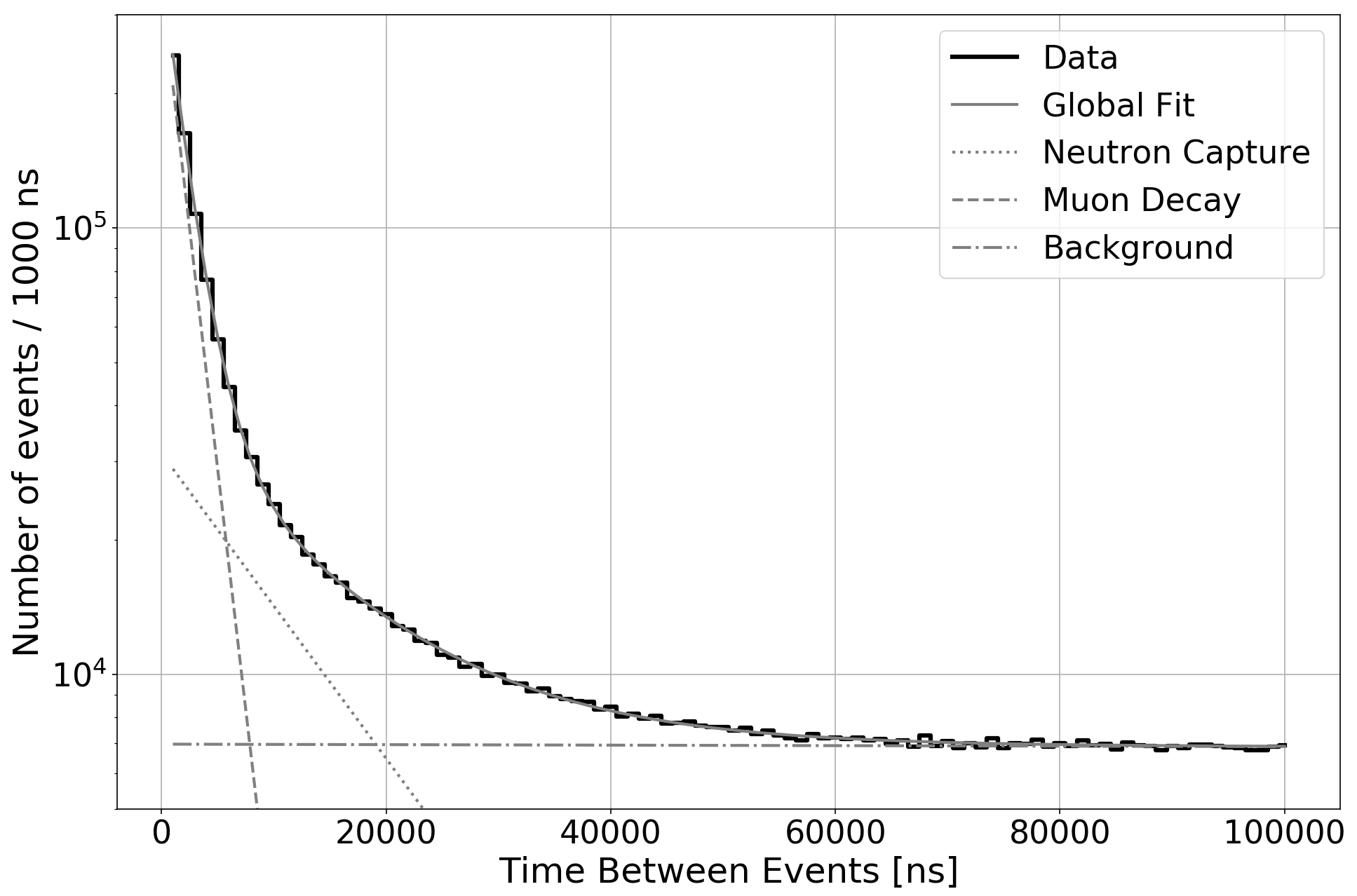}\\
	\caption{Time between events in the Target detector.}
	\label{fig:time_between}
\end{figure}

The commissioning data set has also provided very encouraging results about the ability of the detector to perform the physics program. As previously said, background rejection strategies play the key role to tag genuine neutrino events.
From the commissioning data we obtained the time distribution of time intervals between consecutive events in the Target, shown in figure \ref{fig:time_between}. We have successfully identified in the time range of hundreds of $\mu$s three distinct components: (i) muon decays; (ii) neutron capture in the Gd; (iii) background rate from random coincidences\footnote{The background was fixed from a fit with larger temporal scale. Muon decays and neutron capture were obtained leaving the parameters of two exponential decays free. We should mention that using only one exponential plus background, the resulting time constant is close to the expected for muon decay. The inclusion of a third component (neutron capture hypothesis) improved the fit quality.}. The neutron capture time constant is consistent with the expectations ($\sim$ 15$\mu$s) from the Gd concentration in the Target water.  The fitting result can be seen in table \ref{tab:time_constants}.\\

\begin{table}[ht!]
\centering
\begin{tabular}{c|c|c|}
\cline{2-3}
& \textbf{Time Constant (ns)} & \textbf{Fit Error (ns)} \\ \hline
\multicolumn{1}{|c|}{\textbf{Neutron Capture}} & 12700                       & 80                      \\ \hline
\multicolumn{1}{|c|}{\textbf{Muon decay}}      & 2015                        & 4                       \\ \hline
\multicolumn{1}{|c|}{\textbf{Backgroud}}       & $7.83*10^6$                     & $5*10^4$                   \\ \hline                
\end{tabular}
\caption{Fitting results from data acquired on last commissioning period.}
\label{tab:time_constants}
\end{table}

\begin{figure}
    \includegraphics[scale=0.5]{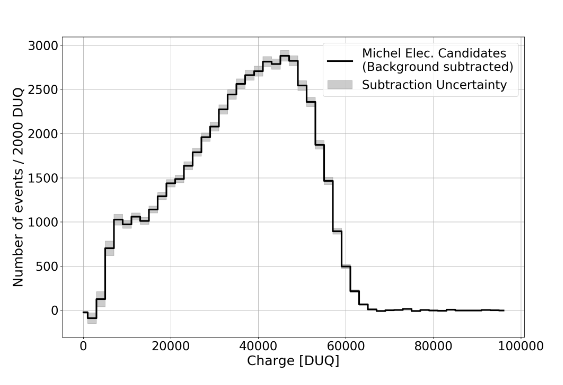} 
  \caption{Michel electrons candidates.}
  \label{fig:michel}
\end{figure}

Additionally, the charge spectrum of Michel electrons\footnote{Strictly speaking, there are Michel \emph{electrons and positrons} in the sample, we referred only to electrons to follow the field jargon.} (ME) candidates is shown in figure \ref{fig:michel}. The lack of the sharp end in the spectrum is due to the response function of the detector. Moreover, the flat region in the low energy range is consistent with the superposition of ME spectrum and the 8 MeV energy deposition of gammas from Gd de-excitation, both convoluted with the detector response function. 

This preliminary analysis from commissioning data shows the good detector performance, by the consistence of time and charge measurements with the expected operational range of the detector. The search for neutrino candidates has been started and is very promising considering the results discussed above. 

\section{Conclusions and perspectives}

\subsection{Experimental results}
The IAEA has a program to induct research in a world-wide scale aiming to identify novel technologies for non-proliferation safeguards. The water Cherenkov detector of the Neutrinos Angra Experiment was successful commissioned. Preliminary data analysis have demonstrated the stability of detector, feasibility of remote operations (key ingredient in the IAEA safeguards program) and ability of the detector to work in the time and energy regime required for antineutrino detection. The Neutrinos Angra Experiment is showing promising results towards the use of an antineutrino detector to monitor nuclear reactors. Special emphasis should be given to the challenging conditions to run the detector at the surface, with no overburden shielding. Up to now, the veto system of the detector and additional techniques for rejection of the huge background show that we are in the right path to start the search for neutrinos candidates. ME analysis has been extensively used to demonstrate all of these above mentioned features. 

\subsection{The future}
We would like bring the attention to the importance to keep the $\nu$-Angra detector running in the Angra nuclear complex, based on some facts:
\begin{enumerate}
    \item The Neutrinos Angra Experiment was totally made by Brazilian scientists and engineers. We have designed, prototyped, built, tested, and commissioned the detector, the FE electronics, and the DAQ boards. The whole R\&D was made in Brazilian labs also in cooperation with local commercial partners, demonstrating the maturity of Brazilian experimental groups to conduct high-level research with autonomy.
    \item We have preliminary results showing that the data has been taken with high quality, and the physics program is close to be completed. The current goal of the experiment is to demonstrate the ability of the detector to count antineutrino events in correlation with the delivered thermal power of the reactor. The future physics program might include measurements of the fractions of nuclear isotopes in the nuclear fuel.
    \item Different technologies can be explored in a second run. Stringent safety rules imposed by the reactor operations put severe restrictions to the materials and equipment allowed inside the power plant. We are in preliminary stages to check the possibility to use water based liquid scintillator (WBLS). WBLS is in total compliance with the safety rules and provides larger energy resolution than water Cherenkov targets. Such an enhancement opens additional physics topics to be explored by the experiment. The compactness of the experiment can also be enhanced by using silicon photomultipliers (SiPMS). The SiPMS can operate easily in the photon counting regime with low voltage power supplies (tens of V), avoiding the extreme care required by the High Voltage (kV) operation of PMTs. The demands on new technologies R\&D can have a significant impact by increasing the knowledge and expertise of local research groups.
    \item The Angra nuclear reactors have shown to be an excellent tool for the development of particle detectors technology and also to perform particle physics research. The experiments $\nu$-Angra  \cite{AngraCommissioning} and CONNIE \cite{connie, connie-phys} are sharing space and running in the neutrino lab. The lab is an adapted freighter container placed close to the reactor dome. We have successfully created a research facility with the cooperation of the power plant operator. Thus, we have created a Latin American research facility that can be used by different groups to carry experimental programs using the reactor as a particle source. This facility has a large potential to foster the growth of Latin American groups doing research using nuclear reactors, due to the closer localization when compared to overseas labs. Moreover, the facility can be very attractive for international researchers, opening the possibility of a healthy exchanging of knowledge and technology.
    
\end{enumerate}

Summarizing, the future of the scientific facility (the neutrino lab) created at the Angra dos Reis power plant is very promising. The two collaborations running neutrinos experiments, $\nu$-Angra and CONNIE, have successfully demonstrated the feasibility to conduct high-level and rather complex experiments in cooperation with the power plant operator. We hope that the neutrino lab in Angra dos Reis can insert Latin America in the world map of research facilities in particle physics.


\end{document}